\documentclass[conference]{IEEEtran}
\IEEEoverridecommandlockouts
\usepackage[hidelinks]{hyperref}
\usepackage{cite}
\usepackage{amsmath,amssymb,amsfonts}
\usepackage{algorithmic}
\usepackage{graphicx}
\usepackage{textcomp}
\usepackage{xcolor}
\def\BibTeX{{\rm B\kern-.05em{\sc i\kern-.025em b}\kern-.08em
    T\kern-.1667em\lower.7ex\hbox{E}\kern-.125emX}}

\usepackage[caption=false,font=footnotesize]{subfig}
\usepackage{algorithmic}
\usepackage{algorithm}

\makeatother

\begin{document}
\title{\title{RIS-Assisted OTFS Communications: Phase Configuration via Received Energy Maximization \thanks{This publication has emanated from research conducted with the financial support of Science Foundation Ireland under Grant numbers SFI/21/US/3757 and SFI/19/FFP/7005(T).} }}
\author{\IEEEauthorblockN{Mohamad H. Dinan and Arman Farhang}\IEEEauthorblockA{\\Department of Electronic and Electrical Engineering, Trinity College
Dublin, Dublin, Ireland \\ Emails: \{mohamad.dinan@tcd.ie, arman.farhang@tcd.ie\}.}}
\maketitle
\begin{abstract}
In this paper, we explore the integration of two revolutionary technologies,
reconfigurable intelligent surfaces (RISs) and orthogonal time frequency
space (OTFS) modulation, to enhance high-speed wireless communications.
We introduce a novel phase shift design algorithm for RIS-assisted
OTFS, optimizing energy reception and channel gain in dynamic environments.
The study evaluates the proposed approach in a downlink scenario,
demonstrating significant performance improvements compared to benchmark
schemes in the literature, particularly in terms of bit error rate
(BER). Our results showcase the potential of RIS to enhance the system's
performance. Specifically, our proposed phase shift design technique
outperforms the benchmark solutions by over 4~dB. Furthermore, even
greater gains can be obtained as the number of RIS elements increases.
\end{abstract}

\begin{IEEEkeywords}
RIS, OTFS modulation, delay-Doppler, 6G.
\end{IEEEkeywords}

\section{Introduction\label{sec:Introduction}}

Reconfigurable intelligent surfaces (RISs) have gained attention due
to increased connectivity demands and the surge in mobile data traffic.
This state-of-the-art technology is set to play a crucial role in
6th-generation wireless networks (6G). RIS is a surface composed of
electromagnetic metamaterial with numerous small, cost-effective,
and energy-efficient reflecting elements \cite{di2020smart}.
These elements can manipulate the scattering and propagation in the
wireless channel by applying a predetermined phase shift to the incoming
wave. Effectively, RIS technology presents a transformative paradigm
shift, that can convert the unpredictable and disruptive propagation
environment into a smart radio setting. This results in an improvement
of received signal quality and provides a revolutionary advancement
in wireless communications \cite{dinan2022ris}.

On the other hand, orthogonal time frequency space (OTFS) modulation
has emerged as a novel solution to address the diverse requirements
of the 5th-generation wireless networks (5G), particularly in scenarios
with high mobility, such as vehicle-to-vehicle communication and high-speed
trains \cite{hadani2017otfs}. Traditional modulation
schemes like orthogonal frequency division multiplexing (OFDM) face
challenges in maintaining effective channel estimation at high speeds.
In contrast, OTFS introduces a revolutionary approach by transforming
the time-varying multipath channel into a two-dimensional domain,
specifically the delay-Doppler (DD) domain \cite{hadani2017otfs,thaj2020rake}.
Such transformation, combined with equalization in the DD domain, ensures
that each transmitted symbol experiences a nearly constant channel
gain. This process involves spreading all information symbols
across both time and frequency dimensions, which leads to the exploitation
of maximum effective diversity \cite{ravi2020diversity}. Hence, OTFS
proves to be a flexible modulation technique that combines features
from both code division multiple access (CDMA) and OFDM \cite{hadani2017otfs}.

The inherent benefits of both RIS and OTFS have inspired researchers
to integrate these cutting-edge technologies to create a robust and
energy-efficient approach, effectively tackling challenges at high-speed wireless communications. Specifically, in \cite{harsha2022risOTFS}
and \cite{vighnesh2023fractional}, RIS-aided OTFS modulation was
proposed to improve the performance of OTFS, and the relative DD domain
input-output relationship was derived. The authors in \cite{muye2022rismilliConf,muye2022rismilli}
investigated a hybrid RIS-aided millimeter wave OTFS system and proposed
a message-passing algorithm to perform channel estimation and data
detection simultaneously. In addition, in \cite{li2023channel} a
novel channel estimation technique was proposed to reduce the training
overhead in an RIS-aided OTFS setup. Inspired by the approach utilized
in RIS-assisted OFDM systems, the researchers in \cite{amit2023risOTFS}
and \cite{thomas2023risOTFS} proposed a phase shift adjustment method
only focusing on the strongest cascaded path. Despite the algorithm's
advantage of low complexity, its focus on the energy of a single path
limits its ability to deliver superior performance, particularly in
situations featuring multiple taps (paths) with close gains. Furthermore,
this technique does not account for the time-varying nature of the
channel.

Most studies investigating RIS have primarily concentrated on single-carrier
scenarios for phase adjustment. Even in OFDM systems, where the entire
bandwidth is considered, there has been a neglect of accounting for
time variations of the channel in high-speed scenarios. Against this
background, we investigate the RIS-assisted OTFS system in this paper
and establish a connection with conventional OTFS without RIS. This
paper considers the complete bandwidth of an OTFS block and assumes
fixed phase shifts of RIS elements throughout an OTFS time frame.
Hence, we introduce a phase shift design algorithm wherein the RIS
is configured to address all paths and considers the dynamic nature
of the channel. The key contributions of this paper are as follows:
\begin{itemize}
\item We study the RIS-assisted OTFS system in a downlink scenario where
the base station (BS) transmits the OTFS signal to a high-speed mobile
terminal (MT) via reflection through an RIS. We find the characteristics
of the cascaded channel that can be treated as a multipath channel
with an increased number of taps. This finding indicates that both the delay time of the BS-RIS link and the Doppler shift of the RIS-MT link contribute to the gain of the cascaded path.
\item We propose a phase shift design to maximize the received energy
at the receiver while enhancing the channel gain. In the proposed
design, we consider the delay-Doppler (DD) channel to effectively
collect the energy of all channel taps within the entire OTFS time
frame to tackle time variations of the channel. Hence, in this paper,
the variations in the channel across both time and frequency domains
contribute to the phase shift adjustment. The proposed algorithm has
low complexity by taking advantage of the sparsity of the DD channel.
\item Finally, we investigate the performance of the system through numerical
analyses and compare the bit error rate (BER) performance results
with those of the benchmark schemes. The results show that the RIS
can significantly enhance the system's performance. This enhancement
follows the results obtained for RIS-assisted single carrier systems.
In addition, we show that the proposed phase shift design substantially
outperforms the benchmark schemes \cite{amit2023risOTFS,thomas2023risOTFS}.
The performance gain over the benchmark schemes is more than 4~dB.
This superiority increases with an increasing number of RIS elements.
\end{itemize}
\emph{Notation:} Boldface lower-case letters denote column vectors,
and boldface upper-case letters denote matrices. $(\cdot)^{\Re}$
and $(\cdot)^{\Im}$ represent the real and imaginary components of
a scalar/vector, respectively. $(\cdot)^{\star}$ signifies the optimum
value of a scalar/vector variable, while $(\cdot)^{*}$ denotes the
complex conjugate. The transpose and Hermitian are denoted by $(\cdot)^{{\rm T}}$
and $(\cdot)^{{\rm H}}$, respectively. $\mathcal{CN}(\mu,\sigma^{2})$
indicates the complex normal distribution with mean $\mu$ and variance
$\sigma^{2}$. For a real/complex scalar $s$, $|s|$ denotes the
absolute value. $[\cdot]_M$ represents modulo $M$ operation. Finally, the set of complex matrices of size $m\times n$
is represented by $\mathbb{C}^{m\times n}$.

\section{System Model\label{sec:System_Model}}

\begin{figure}[t]
\begin{centering}
\includegraphics[scale=0.25]{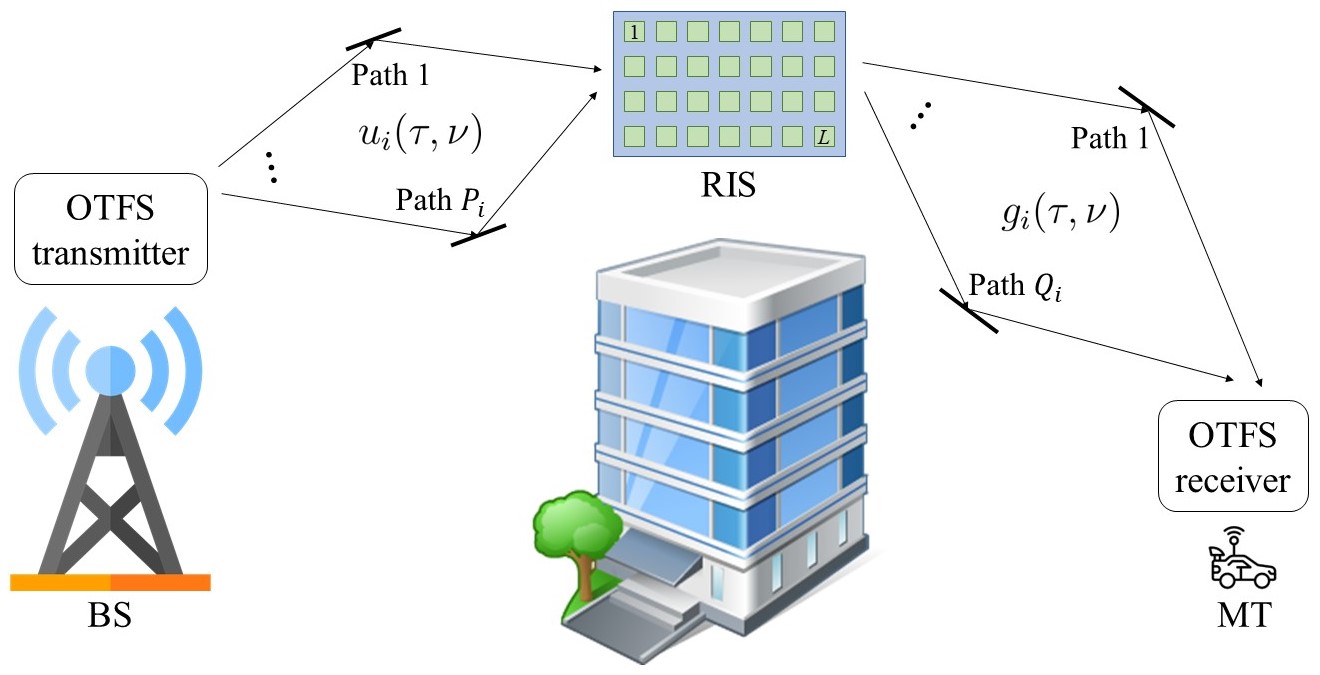}
\par\end{centering}
\caption{An RIS-assisted OTFS setup.\label{fig:Schematic-RIS-OTFS}}
\end{figure}

Fig.~\ref{fig:Schematic-RIS-OTFS} illustrates an RIS-assisted OTFS
system. In this scenario, we examine a downlink communication setup where a BS and a high-speed MT, both equipped with a single antenna, communicate via an RIS. However, the direct link between the transmitter and the receiver is blocked by obstacles. {\color{black}Furthermore, we consider the time-division duplexing (TDD) scenario, while assuming perfect synchronization and knowledge of channel state information (CSI) at the transmitter and receiver.} 
The RIS consists of $L$ reflecting elements whose phase shifts can be adjusted
by the transmitter. The phase shifts vector of the RIS elements is
denoted by $\boldsymbol{\theta}=[\theta_{0},\theta_{1},\dots,\theta_{L-1}]^{{\rm T}}\in\mathbb{C}^{L\times1}$.
We consider a passive RIS with lossless elements, i.e., $\theta_{i}=e^{j\phi_{i}}$
for $i=0,1,\dots,L-1$.

In an OTFS system, the transmit symbols are considered in the delay-Doppler
(DD) domain. Suppose $x[k,l]$ be the data symbol transmitted
in the $k$-th Doppler and $l$-th delay bin where $k=0,1,\dots,N-1$
and $l=0,1,\dots,M-1$. Therefore, the total number of $MN$ symbols
are placed on a uniform DD grid. First, the DD domain data symbols
are transformed to the time-frequency (TF) domain through inverse
symplectic finite Fourier transform (ISFFT) operation, i.e., \vspace{-0.2cm}
\[
X[n,m]=\frac{1}{\sqrt{MN}}\sum_{k=0}^{N-1}\sum_{l=0}^{M-1}x[k,l]e^{j2\pi(\frac{nk}{N}-\frac{ml}{M})},
\]
where $n=0,1,\dots,N-1$ and $m=0,1,\dots,M-1$ are the time and frequency
indices, respectively. The TF samples are uniformly spread in a uniform
plane with a total duration of $NT$ and a total bandwidth of $M\Delta f$.
The time and frequency spacings between the samples are $\Delta\tau=T/M$
and $\Delta f=1/T$, respectively. Then, the Heisenberg transform
\cite{hadani2017otfs} is applied to the TF signal to obtain the time
domain transmit signal $s(t)$.

Next, let the DD domain BS-RIS and RIS-MT channels for each RIS element
be, respectively, represented as \vspace{-0.2cm}
\begin{equation}
u_{i}(\tau,\nu)=\sum_{p=0}^{P_{i}-1}u_{i,p}\delta(\tau-\tau_{i,p}^{(u)})\delta(\nu-\nu_{i,p}^{(u)}),\ i=0,1,\dots,L-1,\label{eq:channel_u}
\end{equation}
\begin{equation}
g_{i}(\tau,\nu)=\sum_{q=0}^{Q_{i}-1}g_{i,q}\delta(\tau-\tau_{i,q}^{(g)})\delta(\nu-\nu_{i,q}^{(g)}),\ i=0,1,\dots,L-1,\label{eq:channel_g}
\end{equation}
where $P_{i}$ is the number of paths between the BS and RIS element
$i$, and $u_{i,p}$, $\tau_{i,p}^{(u)}$ and $\nu_{i,p}^{(u)}$ are
the path gain, delay time, and Doppler shift, respectively, corresponding
to path $p=0,1,\dots,P_{i}-1$; while, $Q_{i}$ is the number of paths
between the $i$-th RIS element and MT, and $g_{i,q}$, $\tau_{i,q}^{(g)}$
and $\nu_{i,q}^{(g)}$ denote the path gain, delay time, and Doppler
shift, respectively, associated with path $q=0,1,\dots,Q_{i}-1$.
We consider the multipath channel gains $u_{i,p}$ and $g_{i,q}$,
respectively, distributed as $\mathcal{CN}(0,\sigma_{u_{i,p}}^{2})$
and $\mathcal{CN}(0,\sigma_{g_{i,q}}^{2})$, with $\sum_{p=0}^{P_{i}-1}\sigma_{u_{i,p}}^{2}=1$,
and $\sum_{q=0}^{Q_{i}-1}\sigma_{g_{i,q}}^{2}=1$, for all $i=0,1,\dots,L-1$.

The time domain signal travels through the BS-RIS channel. Hence,
the received signal at RIS element $i$ is given by 
\begin{equation}
y_{i}(t)=\sum_{p=0}^{P_{i}-1}u_{i,p}e^{j2\pi\nu_{i,p}^{(u)}(t-\tau_{i,p}^{(u)})}s(t-\tau_{i,p}^{(u)}),\ i=0,1,\dots,L-1.\label{eq:receive_y}
\end{equation}
Then, the reflected signal from RIS element $i$ is multiplied by
coefficient $\theta_{i}$. The resulting signal is then passed through
the RIS-MT channel. Finally, the received signal at the MT is represented
as \vspace{-0.2cm}
\begin{align}
z(t)= & \sum_{i=0}^{L-1}\theta_{i}\sum_{q=0}^{Q_{i}-1}g_{i,q}e^{j2\pi\nu_{i,q}^{(g)}(t-\tau_{i,q}^{(g)})}y_{i}(t-\tau_{i,q}^{(g)})+w(t)\nonumber \\
%= & \sum_{i=0}^{L-1}\theta_{i}\sum_{q=0}^{Q_{i}-1}g_{i,q}e^{j2\pi\nu_{i,q}^{(g)}(t-\tau_{i,q}^{(g)})}\times\nonumber \\
% & \sum_{p=0}^{P_{i}-1}u_{i,p}e^{j2\pi\nu_{i,p}^{(u)}(t-\tau_{i,p}^{(u)}-\tau_{i,q}^{(g)})}s(t-\tau_{i,p}^{(u)}-\tau_{i,q}^{(g)})+w(t)\nonumber \\
= & \sum_{i=0}^{L-1}\theta_{i}\sum_{q=0}^{Q_{i}-1}\sum_{p=0}^{P_{i}-1}g_{i,q}u_{i,p}e^{j2\pi\nu_{i,q}^{(g)}\tau_{i,p}^{(u)}}\times\nonumber \\
 & e^{j2\pi(\nu_{i,p}^{(u)}+\nu_{i,q}^{(g)})(t-(\tau_{i,p}^{(u)}+\tau_{i,q}^{(g)}))}s(t-(\tau_{i,p}^{(u)}+\tau_{i,q}^{(g)}))+w(t),\label{eq:receive_z}
\end{align}
where $w(t)$ is the additive white Gaussian noise (AWGN). By examining
(\ref{eq:receive_z}), we note that the cascaded RIS-assisted channel
resembles a multipath channel consisting of total $R_{i}=P_{i}Q_{i}$
paths with the DD domain given by \vspace{-0.2cm}
\begin{align}
h_{i}(\tau,\nu)= & \sum_{q=0}^{Q_{i}-1}\sum_{p=0}^{P_{i}-1}h_{i,pq}\delta(\tau-\tau_{i,pq})\delta(\nu-\nu_{i,pq}),\nonumber \\
 & i=0,1,\dots,L-1,\label{eq:totalchannel_h}
\end{align}
where $h_{i,pq}=g_{i,q}u_{i,p}e^{j2\pi\nu_{i,q}^{(g)}\tau_{i,p}^{(u)}}$,
$\tau_{i,pq}=\tau_{i,p}^{(u)}+\tau_{i,q}^{(g)}$, and $\nu_{i,pq}=\nu_{i,p}^{(u)}+\nu_{i,q}^{(g)}$.
It is important to note that the cumulative delay times and Doppler
shifts of cascaded paths are the summation of individual delay times
and Doppler shifts. Meanwhile, the combined gain of successive paths
is determined by multiplying individual gains, incorporating a phase
shift that is related to the Doppler shift of the RIS-MT link and
the delay time of the BS-RIS link. With this realization, we can leverage
the results obtained for conventional OTFS systems without RIS. Hence, we can
rewrite the received signal as
\begin{equation}
z(t)=\sum_{i=0}^{L-1}\theta_{i}\sum_{q=0}^{Q_{i}-1}\sum_{p=0}^{P_{i}-1}h_{i,pq}e^{j2\pi\nu_{i,pq}(t-\tau_{i,pq})}s(t-\tau_{i,pq})+w(t).\label{eq:receive_z2}
\end{equation}

Using the Wigner transform \cite{hadani2017otfs} at the receiver
the TF signal is obtained. Assuming bi-orthogonal transmit and receive
pulses, then, by sampling the received TF signal at $t=nT$ and $f=m\Delta f$,
the discrete TF signal $Z[n,m]$ can be derived as \cite{raviteja2018interference} \vspace{-0.3cm}
\begin{equation}
Z[n,m]=\sum_{i=0}^{L-1}\theta_{i}H_{i}[n,m]X[n,m]+W[n,m],\label{eq:Z=00005Bm,n=00005D}
\end{equation}
where $W[n,m]$ is the discrete noise signal at the TF bin independent
and identically distributed (i.i.d.) as $\mathcal{CN}(0,\sigma_{0}^{2})$
for all $n$ and $m$. $H_{i}[n,m]$ is the TF cascaded channel associated
with RIS element $i$ that is given by \cite{raviteja2018interference}
\begin{align*}
H_{i}[n,m]= & \int_{\tau}\int_{\nu}h_{i}(\tau,\nu)e^{j2\pi\nu nT}e^{-j2\pi(\nu+m\Delta f)\tau}\mathrm{d}\nu\mathrm{d}\tau.
\end{align*}
Next, the TF domain signal is transformed to the DD domain by a symplectic
finite Fourier transform (SFFT) operation, i.e., 
\[
z[k,l]=\frac{1}{\sqrt{MN}}\sum_{n=0}^{N-1}\sum_{m=0}^{M-1}Z[n,m]e^{-j2\pi(\frac{nk}{N}-\frac{ml}{M})}.
\]
Hence, the DD domain input-output relationship can be expressed as \cite[Eq. (20)]{raviteja2018interference} %\vspace{-0.3cm}
\begin{align}
z[k,l]\approx & \sum_{i=0}^{L-1}\theta_{i}\sum_{q=0}^{Q_{i}-1}\sum_{p=0}^{P_{i}-1}\sum_{n'=-N'}^{N'}\left(\frac{e^{-j2\pi(-n'-k'_{i,pq})}-1}{Ne^{-j\frac{2\pi}{N}(-n'-k'_{i,pq})}-N}\right) \nonumber\\
 & \times h_{i,pq}e^{-j2\pi\nu_{i,pq}\tau_{i,pq}}\nonumber \\
 & \times x[[k-k_{i,pq}+n']_{N},[l-l_{i,pq}]_{M}]+w[k,l],\label{eq:z=00005Bk,l=00005D}
\end{align}
where $N'\ll N$, ${k_{i,pq}}+{k'_{i,pq}}=\nu_{i,pq}{NT}$, and ${l_{i,pq}}=\tau_{i,pq}{M\Delta f}$. $k_{i,pq}$ and $k'_{i,pq}$ represent the integer and fractional parts of the normalized Doppler shifts $\nu_{i,pq}NT$, respectively. Here, we consider integer delay shifts ${l_{i,pq}}$, which is a valid assumption in typical wideband systems.

Then, (\ref{eq:z=00005Bk,l=00005D}) can be represented in matrix
form as %\vspace{-0.2cm}
\begin{equation}
\mathbf{z}=\sum_{i=0}^{L-1}\theta_{i}\mathbf{H}_{i}\mathbf{x}+\mathbf{w},\label{eq:matrix_form}
\end{equation}
where $\mathbf{z},\mathbf{x}\in\mathbb{C}^{MN\times1}$ are the vectored
forms of $z[k,l]$ and $x[k,l]$, respectively, such that $z_{k+Nl}=z[k,l]$
and $x_{k+Nl}=x[k,l]$. $\mathbf{H}_{i}\in\mathbb{C}^{MN\times MN}$
is the cascaded DD domain channel matrix, and $\mathbf{w}\in\mathbb{C}^{MN\times1}$
is the i.i.d. noise vector whose elements are with the same distribution
as $W[n,m]$. 

\section{Proposed Phase Shifts Design\label{sec:phase_design}}

In this section, we aim to adjust the phase shifts of the RIS elements
to improve the performance of the OTFS. It is noteworthy that the
majority of studies addressing RIS have predominantly focused on single-carrier
scenarios for phase adjustment. Even in OFDM systems, where the entire
bandwidth is taken into account, there has been a lack of consideration
for channel aging in mobile scenarios. In contrast, this paper addresses
the full bandwidth of an OTFS block and assumes fixed phase shifts
of RIS elements during an OTFS time frame. Consequently, the channel
variation in both time and frequency domains contribute to the phase
adjustment.\footnote{It is worth noting that, similar to previous studies \cite{harsha2022risOTFS,amit2023risOTFS,thomas2023risOTFS},
we assume that the parameters of the DD domain channel remain relatively
unchanged over a series of consecutive OTFS blocks while acknowledging
variations from sample to sample.} In light of this, taking into account the channel knowledge available at the transmitter, we aim to enhance the performance of the OTFS system by considering the signal-to-noise ratio (SNR) at the receiver, incorporating the Delay-Doppler (DD) channel. To this end, we propose an RIS phase adjustment
technique that maximizes the received signal energy at the MT. Hence,
the optimization problem can be defined as %\vspace{-0.2cm}
\begin{align}
\mathrm{P1}:\ \max_{\boldsymbol{\theta}} & \ \ G(\boldsymbol{\theta})=\left\Vert \sum_{i=0}^{L-1}\theta_{i}\mathbf{H}_{i}\right\Vert _{F}^{2}\nonumber \\
\ \ \ \mathrm{s.t.}\ \  & \ \ \mathcal{S}:\left|\theta_{i}\right|=1,\ i=0,1,\dots,L-1,\label{eq:P1}
\end{align}
where $\left\Vert \cdot\right\Vert _{F}$ computes the Frobenius norm
of a matrix. The solution for $\mathrm{P1}$ is not straightforward
as it is a non-convex optimization problem. Hence, we adopt a gradient
method to find an approximate solution to $\mathrm{P1}$ \cite{JMLR:v11:journee10a}.
Algorithm~\ref{alg:gradient_scheme} shows this gradient-based procedure.
The gradient of $G(\boldsymbol{\theta})$ needs to be computed to
implement this algorithm. For this reason, we rewrite $G(\boldsymbol{\theta})$
as %\vspace{-0.3cm}
\[
G(\boldsymbol{\theta})=\left\Vert \sum_{i=0}^{L-1}\theta_{i}\bar{\mathbf{h}}_{i}\right\Vert _{2}^{2},
\]
where $\bar{\mathbf{h}}_{i}=\mathrm{vec}(\mathbf{H}_{i})$, and $\left\Vert \cdot\right\Vert _{2}$
is the Euclidean norm operator. To calculate the gradient of $G(\boldsymbol{\theta})$,
we find its partial derivatives with respect to the complex conjugate
of $\theta_{i}$, i.e., $\frac{\partial G(\boldsymbol{\theta})}{\partial\theta_{i}^{*}}$
based on Wirtinger calculus \cite{fischer2005precoding}. Hence, the
gradient of $G(\boldsymbol{\theta})$ can be expressed as %\vspace{-0.2cm}
\[
\nabla G(\boldsymbol{\theta})=\left[\frac{\partial G(\boldsymbol{\theta})}{\partial\theta_{0}^{*}},\frac{\partial G(\boldsymbol{\theta})}{\partial\theta_{1}^{*}},\dots,\frac{\partial G(\boldsymbol{\theta})}{\partial\theta_{L-1}^{*}}\right]^{{\rm T}},
\]
\begin{equation}
\frac{\partial G(\boldsymbol{\theta})}{\partial\theta_{i}^{*}}=\sum_{\ell=0}^{L-1}\theta_{\ell}\bar{\mathbf{h}}_{i}^{{\rm H}}\bar{\mathbf{h}}_{\ell},\ i=0,1,\dots,L-1.\label{eq:gradient}
\end{equation}
Next, the optimization problem at line~6 of Algorithm~\ref{alg:gradient_scheme},
which needs to be solved in each iteration, is given by 
\begin{align}
\mathrm{P2}:\ \max_{\bar{\boldsymbol{\theta}}} & \ \ \left(\nabla G(\boldsymbol{\theta}^{(j)}){}^{{\rm H}}\bar{\boldsymbol{\theta}}\right)^{\mathcal{\Re}}=\sum_{i=0}^{L-1}\left(\bar{\theta}_{i}^{\Re}\gamma_{i}^{\mathcal{\Re}}+\bar{\theta}_{i}^{\mathcal{\Im}}\gamma_{i}^{\Im}\right)\nonumber \\
\ \ \ \text{s.t.}\ \  & \ \ \mathcal{S}:\left|\bar{\theta}_{i}\right|=1,\ i=0,1,\dots,L-1,\nonumber \\
\text{where} & \ \ \boldsymbol{\gamma}\triangleq\nabla G(\boldsymbol{\theta}^{(j)}),\ \text{for iteration }j.\label{eq:P2}
\end{align}
It is worth noting that as the main objective function in $\mathrm{P1}$
has a positive real value, we are interested in the real direction
of the gradient. Hence, we consider the maximization of the real part
in $\mathrm{P2}$. Problem $\mathrm{P2}$ is still a non-convex problem
due to the constraint; however, it can be transformed into a convex
problem through relaxation, i.e., using the constraint $\left|\bar{\theta}_{i}\right|\leq1$
for $i=0,1,\dots,L-1$. Next, it can be proved that the optimal solution
for the relaxed problem which is convex with the updated constraint,
is given by 
\begin{equation}
\bar{\theta}_{i}^{\star}=\frac{\gamma_{i}}{\left|\gamma_{i}\right|},\ i=0,1,\dots,L-1;\label{eq:optimal_theta}
\end{equation}
that is $\angle\bar{\theta}_{i}^{\star}=\angle\gamma_{i}$ and $\left|\bar{\theta}_{i}^{\star}\right|=1$.
Therefore, problem $\mathrm{P1}$ can be solved by implementing the
procedure presented in Algorithm~\ref{alg:proposed_sol}. {\color{black}We establish a stopping criterion for the algorithm by setting a threshold condition  $\frac{G(\boldsymbol{\theta}^{(j)})-G(\boldsymbol{\theta}^{(j-1)})}{G(\boldsymbol{\theta}^{(j-1)})}<\epsilon$, where $\epsilon$ is a design parameter.}

\subsubsection*{Complexity Analysis}It can be realized that the main complexity of this algorithm lies
at line~6, where the gradient needs to be computed at $\boldsymbol{\theta}^{(j)}$.
Note that \{${\mathbf{H}_{i}}$\} is a block-circulant (BC) matrix.
Therefore, \{${\bar{\mathbf{h}}_{i}}$\} can be obtained by considering only the first $M$ columns of \{${\mathbf{H}_{i}}$\}. As a result, initially, the complexity seems to be in the order of $L^{2}MN^{2}$,
i.e., $\mathcal{O}\{\xi L^{2}MN^{2}\}$, where $\xi$ is the number of iterations. However, considering the sparsity of \{${\mathbf{H}}_{i}$\}, the number of non-zero elements in \{$\bar{\mathbf{h}}_{i}$\}
is far less than its length, verifying the low complexity of this
algorithm. In other words, the complexity of this process is $\mathcal{O}\{\xi L^{2}\mathcal{B}\}$,
where $\mathcal{B}\leq R\ll MN^{2}$, and $R\approx\alpha M$
is the maximum number of non-zero elements in $\{\bar{\mathbf{h}}_{i}$\},
while $\alpha$ is the number of nonzero elements in each row of ${\mathbf{H}_{i}}$.
The equality occurs when all channels experience the same number of
taps; in this case, the coincidence of nonzero elements in the vectorized
channels $\{\bar{\mathbf{h}}_{i}$\} is the same for all $i=0,1,\dots,L-1$.

\begin{algorithm}[h!]
\begin{algorithmic}[1]

\STATE \textbf{input:} $\mathbf{H}_i,\ i=0,1,\dots,L-1$ 
\STATE \textbf{output:} $\boldsymbol{\theta}^\star$; approximate solution of $\mathrm{P1}$ 
\STATE \textbf{initialize:} $\boldsymbol{\theta} = \boldsymbol{\theta}^{(0)} \in \mathcal{S}$ 
\STATE $j \leftarrow 0$ 
\REPEAT 
\STATE $\boldsymbol{\theta}^{(j+1)}=\arg\max_{\bar{\boldsymbol{\theta}}}\{{G}(\boldsymbol{\theta}^{(j)})+\nabla{G}(\boldsymbol{\theta}^{(j)})^{\rm{H}}(\bar{\boldsymbol{\theta}}-$
\STATE \hspace{1.4cm}$\boldsymbol{\theta}^{(j)})|\bar{\boldsymbol{\theta}}\in\mathcal{S}\}$ 
\STATE $j \leftarrow j+1$ 
\UNTIL \text{``a specific condition is met''} 
\STATE $\boldsymbol{\theta}^\star=\boldsymbol{\theta}^{(j)}$

\end{algorithmic} 

\caption{Gradient procedure for maximizing a convex function.\label{alg:gradient_scheme}}
\end{algorithm}
\vspace{-0.3cm}
\begin{algorithm}[h!]
\begin{algorithmic}[1]

\STATE \textbf{input:} $\epsilon,\ \bar{\mathbf{h}}_i,\ i=0,1,\dots,L-1$ 
\STATE \textbf{output:} $\boldsymbol{\theta}^\star$; approximate solution of $\mathrm{P1}$ 
\STATE \textbf{initialize:} $\boldsymbol{\theta} = \boldsymbol{\theta}^{(0)} \in \mathcal{S}$ 
\STATE $j \leftarrow 0$ 
\REPEAT
\STATE calculate $\gamma_{i}=\sum_{\ell=0}^{L-1}\theta_{\ell}^{(j)}\bar{\mathbf{h}}_{i}^{\rm{H}}\bar{\mathbf{h}}_{\ell},\ i=0,1,\dots,L-1$
\STATE $\theta_{i}^{(j+1)}=\frac{\gamma_{i}}{\left|\gamma_{i}\right|},\ i=0,1,\dots,L-1$
\STATE $j \leftarrow j+1$ 
\UNTIL $\frac{G(\boldsymbol{\theta}^{(j)})-G(\boldsymbol{\theta}^{(j-1)})}{G(\boldsymbol{\theta}^{(j-1)})}<\epsilon$ 
\STATE $\boldsymbol{\theta}^\star=\boldsymbol{\theta}^{(j)}$

\end{algorithmic} 

\caption{Proposed solution for adjusting the phase shifts of RIS elements.\label{alg:proposed_sol}}
\end{algorithm}

\section{Numerical Results\label{sec:Numerical_Results}}

In this section, we demonstrate the performance of the RIS-assisted
OTFS system through numerical analyses. First, we present the performance
of the system across various scenarios, emphasizing the efficacy of
the proposed phase shift design and optimization algorithm. Next,
we conduct a comparison of the BER performance between the RIS-assisted
OTFS system with the proposed phase shift design and benchmark schemes
\cite{amit2023risOTFS,thomas2023risOTFS}. These benchmarks consist
of prominent RIS-assisted OTFS systems with the same configuration
as this paper. Considering their notable improvements over OFDM systems,
we find the comparison with OFDM scenarios unnecessary in this context.
Without loss of generality, in our simulations, we consider the scenario
where both the BS and RIS are fixed at high elevations. Hence, it
is assumed that all multipath channels in the BS-RIS link experience
the same number of paths with the same delay time and without Doppler
effects; while the MT is relatively far from the RIS, such that the
multipath channels between them experience the same number of paths
with the same delay, but different Doppler taps. In addition, at the
receiver, we deploy the minimum mean square error (MMSE) for all scenarios,
while assuming the CSI available at the receiver.

\begin{figure}[t]
\begin{centering}
\includegraphics[scale=0.47]{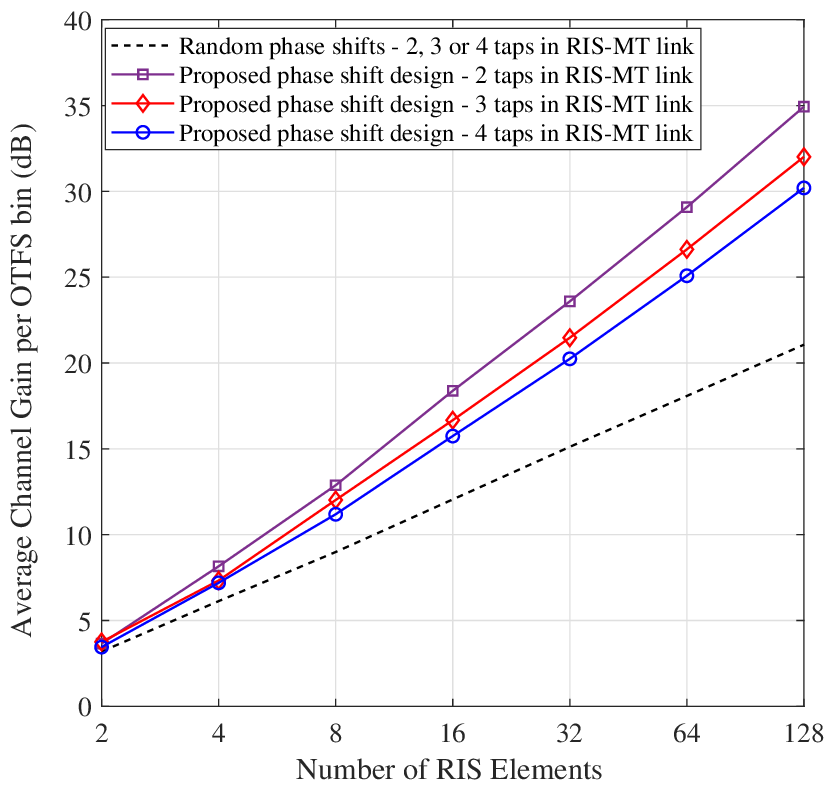}
\par\end{centering}
\caption{Effect of increasing number of RIS elements on the average channel gain
in RIS-assisted OTFS systems; here, $N=16$ and $M=32$.\label{fig:received_energy}}

\end{figure}

\begin{figure}[t]
\begin{centering}
\includegraphics[scale=0.47]{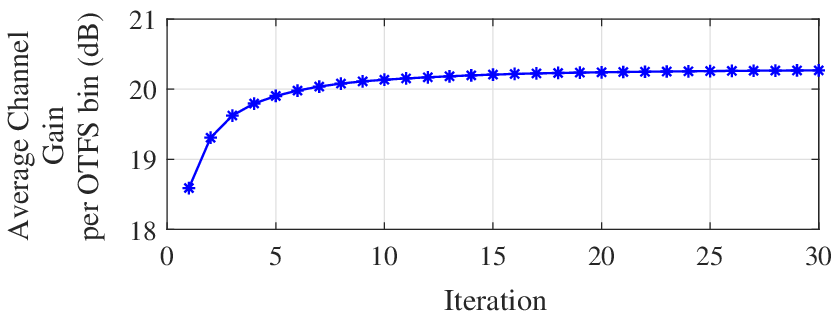}
\par\end{centering}
\caption{Convergence rate of the optimization algorithm; here, $L=32$, $N=16$
and $M=32$.\label{fig:convergence}}
\end{figure}

Fig.~\ref{fig:received_energy} shows the average channel gain in
an RIS-assisted OTFS system with $N=16$ Doppler bins and $M=32$
subcarriers, where the phase shifts of the RIS elements are designed
based on Algorithm~\ref{alg:proposed_sol}. In addition, we consider
four delay taps in the BS-RIS link with \emph{equal average gain power},
while we present the results for various numbers of delay and Doppler taps in the RIS-MT link that are randomly selected with \emph{equal
average gain power}. Certainly, increasing the number of RIS elements
noticeably improves the effective gain of the RIS-assisted channel,
specifically when the phase shifts are properly adjusted. It can be
observed that the RIS with the proposed phase shift design remarkably
enhances the performance of the system and verifies the strength
of the optimization algorithm. For example, the system with optimized
phase shifts improves the performance by approximately 10~dB compared
with the system with random phase shifts, in the scenario with four
taps. This superiority increases for scenarios with less number of
taps. In addition, the RIS with random phase shifts can enhance the
channel gain by 3~dB with increasing the number of RIS elements by
a factor of 2. Meanwhile, by the proposed phase design, the channel
gain can be improved by approximately 6~dB, where the number of RIS
elements is relatively large. These results are in accordance with
the results obtained in RIS-assisted schemes for single carrier systems
in time-invariant channels \cite{dinan2023ris}. Additionally,
it is noted that when employing a random phase configuration, the
received energy remains independent of the number of taps. However,
by implementing adjusted phase shifts, further enhancement is achieved
as the number of taps decreases. This is because the RIS can more
efficiently target a smaller number of taps (paths). It is noted that
increasing bandwidth leads to a higher number of taps. According to
this observation, the efficacy of the RIS is subsequently reduced.
Moreover, to validate the efficacy of the proposed algorithm, we illustrate
its convergence rate in Fig.~\ref{fig:convergence}. The graph demonstrates
that the algorithm's output approaches approximately 97\% of the optimum
value after just 10 iterations, confirming the rapid convergence of
the proposed algorithm.

\begin{figure}[t]
\begin{centering}
\includegraphics[scale=0.47]{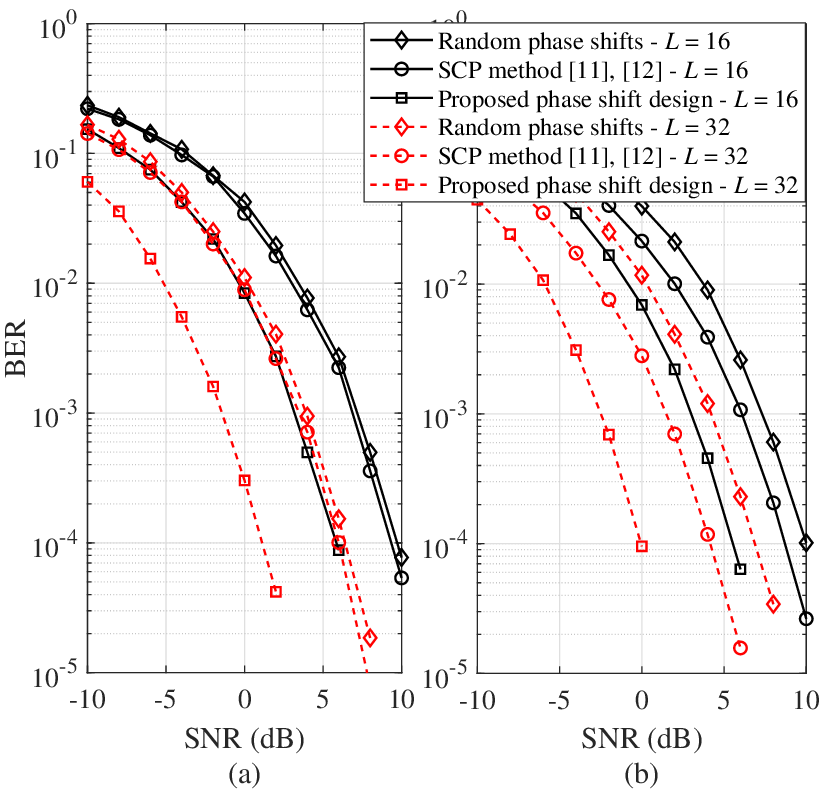}
\par\end{centering}
\caption{BER performance for $N=16$, and $M=32$,
(a) channel with equal average tap gains, and (b) channel with unequal
average tap gains.\label{fig:ber}}
\end{figure}

Next, in Fig.~\ref{fig:ber}, we illustrate the BER performance of
the RIS-assisted OTFS system and compare the results with that of
the benchmark schemes. For benchmark, we consider RIS-assisted OTFS
systems in \cite{amit2023risOTFS,thomas2023risOTFS}, where only the
strongest cascaded path (SCP) was considered for phase shift adjustment.
In all scenarios, we use channels with four taps in both links and
consider the system with $N=16$, $M=32$, and 4-QAM modulation. In
addition, the maximum number of iterations and $\epsilon$ are set to be 15 and $10^{-4}$, respectively, for the
proposed algorithm to design the phase shifts. Fig.~\ref{fig:ber}(a)
shows the BER results in a channel with \emph{equal average tap gains}.
We observe that our proposed phase shift design significantly enhances
the performance of the system. In contrast, adjusting the phase shifts
based on the SCP method does not result in a considerable improvement
over random phase shifts. To have a fair comparison with the SCP method
which considers a single strong tap, in Fig.~\ref{fig:ber}(b), we
consider the channel with \emph{unequal average tap gains} for both
BS-RIS and RIS-MT links, such that 70 percent of the channel gain
in each link is dedicated to a single tap. By observing Fig.~\ref{fig:ber}(b),
we realize that the SCP method can moderately improve the BER performance,
however, the proposed solution can leverage the enhancement more significantly.
For instance, when our proposed technique is utilized, approximately
4~dB and 7~dB performance gains are achieved, respectively, compared
with the SCP method in \cite{amit2023risOTFS,thomas2023risOTFS},
and random phase shifts at the BER of $10^{-4}$, where 32 RIS elements
are deployed. This superiority over the benchmarks is due to the fact
that the SCP method only targets a single tap; hence, the RIS is unable
to efficiently reflect the energy of the other taps toward the receiver.
Although SCP uses a low-complexity algorithm, it cannot provide superior performance and presents even a small improvement if there
exist multiple taps with the same energy. In contrast, the proposed phase shift design collects the energy of all available taps and considers
the channel variation caused by Doppler effects. In addition, it can
be observed that increasing the number of RIS elements from 16 to
32, enhances the performance of the system with random phase shifts
by 3~dB and the system using the SCP method by 4.3~dB, while the
enhancement is around 6~dB for our proposed technique. This reveals
that the superiority over the benchmark schemes dramatically increases
with increasing the number of RIS elements.

\begin{figure}[t]
\begin{centering}
\includegraphics[scale=0.47]{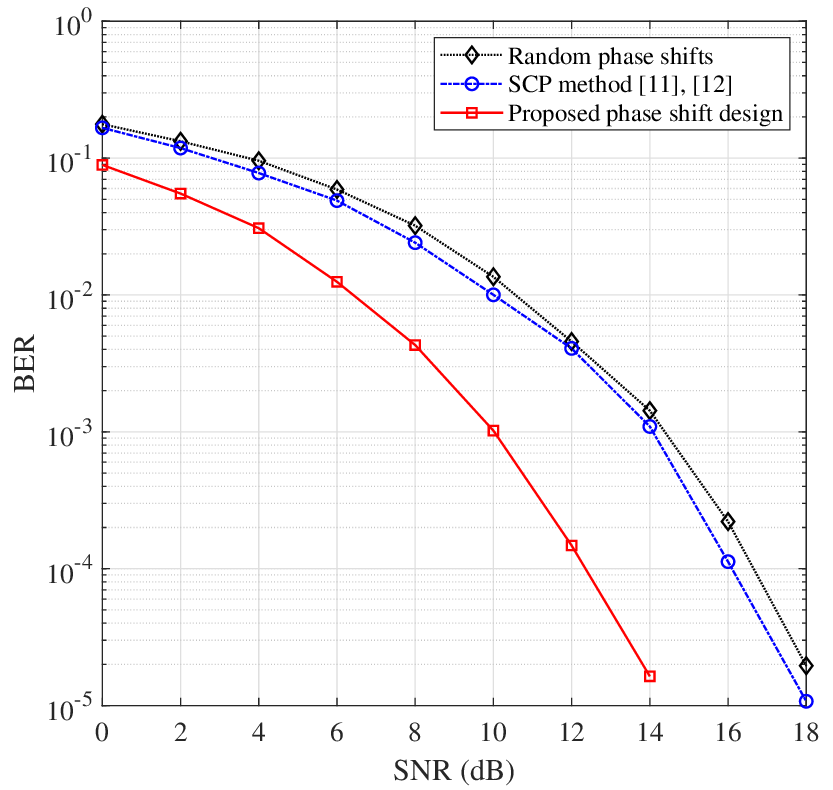}
\par\end{centering}
\caption{BER performance of RIS-assisted OTFS systems in the TDL-C channel;
here, $N=16$, $M=128$, and 4-QAM modulation is used.\label{fig:ber_TDL}}
\end{figure} 

Finally, to showcase the performance in a standard channel scenario,
in Fig.~\ref{fig:ber_TDL} we present the BER performance results
in the tapped delayed line (TDL) channel model, denoted by TDL-C \cite{etsiTR36913},
provided by 3rd generation partnership project (3GPP). In this scenario,
we consider a system with $N=16$ Doppler bins, $M=128$ frequency
subcarriers, and $L=16$ RIS elements, where the carrier frequency
and subcarrier spacing are 4~GHz and 15~kHz, respectively. The MT
speed is assumed to be 500~km/h. As expected, we observe the same
trend as seen in Fig.~\ref{fig:ber}(a). This is because, with the
aforementioned configuration, two taps possess approximately equal
average energy, and the SCP method can only target one of them. That
is, the SCP method cannot efficiently improve the performance, while
the system with the proposed phase shift design outperforms the SCP
method by approximately 4~dB at the BER of $10^{-4}$. 

\section{Conclusion\label{sec:Conclusion}}

This paper investigated the integration of RIS with OTFS modulation
to enhance high-speed wireless communications. We presented the characterization
of the cascaded channel and provided the input-output relation considering
the cascaded delay-Doppler channel response. A phase shift design
algorithm was proposed which maximizes the energy at the receiver
by collecting the energy of all taps in the whole OTFS time frame while
considering the time-varying property of the channel. Moreover, comprehensive
numerical analyses showcase the system's performance improvements.
Results indicate that the proposed approach significantly enhances
the system's performance compared to benchmark schemes, particularly
achieving more than 4~dB gain over them. The findings underscore
the potential of this integrated approach for advancing the capabilities
of wireless communication systems, particularly in high-speed and
dynamic environments. 

\bibliographystyle{ieeetr}
\bibliography{ref}

\begin{thebibliography}{10}

\bibitem{di2020smart}
M.~Di~Renzo {\em et~al.}, ``Smart radio environments empowered by reconfigurable intelligent surfaces: {H}ow it works, state of research, and the road ahead,'' {\em IEEE J. Sel. Areas Commun.}, vol.~38, pp.~2450--2525, Nov. 2020.

\bibitem{dinan2022ris}
M.~H. Dinan, N.~S. Perovi\'c, and M.~F. Flanagan, ``{RIS}-assisted receive quadrature space-shift keying: {A} new paradigm and performance analysis,'' {\em IEEE Trans. Commun.}, vol.~70, pp.~6874--6889, Oct. 2022.

\bibitem{hadani2017otfs}
R.~Hadani, S.~Rakib, M.~Tsatsanis, A.~Monk, A.~J. Goldsmith, A.~F. Molisch, and R.~Calderbank, ``Orthogonal time frequency space modulation,'' in {\em IEEE Wireless Commun. Netw. Conf. (WCNC)}, pp.~1--6, 2017.

\bibitem{thaj2020rake}
T.~Thaj and E.~Viterbo, ``Low complexity iterative rake decision feedback equalizer for zero-padded {OTFS} systems,'' {\em IEEE Trans. Veh. Technol.}, vol.~69, pp.~15606--15622, Dec. 2020.

\bibitem{ravi2020diversity}
P.~Raviteja, Y.~Hong, E.~Viterbo, and E.~Biglieri, ``Effective diversity of {OTFS} modulation,'' {\em IEEE Wireless Commun. Lett.}, vol.~9, pp.~249--253, Feb. 2020.

\bibitem{harsha2022risOTFS}
G.~Harshavardhan, V.~S. Bhat, and A.~Chockalingam, ``{RIS}-aided {OTFS} modulation in high-{D}oppler channels,'' in {\em IEEE 33rd Annu. Int. Symp. Pers., Indoor, Mobile Radio Commun. (PIMRC)}, pp.~409--415, 2022.

\bibitem{vighnesh2023fractional}
V.~S. Bhat, G.~Harshavardhan, and A.~Chockalingam, ``Input-output relation and performance of {RIS}-aided {OTFS} with fractional delay-{D}oppler,'' {\em IEEE Commun. Lett.}, vol.~27, pp.~337--341, Jan. 2023.

\bibitem{muye2022rismilliConf}
M.~Li, S.~Zhang, Y.~Ge, F.~Gao, and P.~Fan, ``A novel transmission strategy for hybrid {RIS} aided millimeter wave {OTFS} systems,'' in {\em 14th Int. Conf. Wireless Commun. Signal Process.}, pp.~1058--1063, 2022.

\bibitem{muye2022rismilli}
M.~Li, S.~Zhang, Y.~Ge, F.~Gao, and P.~Fan, ``Joint channel estimation and data detection for hybrid {RIS} aided millimeter wave {OTFS} systems,'' {\em IEEE Trans. Commun.}, vol.~70, pp.~6832--6848, Oct. 2022.

\bibitem{li2023channel}
Z.~Li, W.~Yuan, B.~Li, J.~Wu, C.~You, and F.~Meng, ``Reconfigurable-intelligent-surface-aided {OTFS}: {T}ransmission scheme and channel estimation,'' {\em IEEE Internet Things J.}, vol.~10, pp.~19518--19532, Nov. 2023.

\bibitem{amit2023risOTFS}
A.~S. Bora, K.~T. Phan, and Y.~Hong, ``{IRS}-assisted high mobility communications using {OTFS} modulation,'' {\em IEEE Wireless Commun. Lett.}, vol.~12, pp.~376--380, Feb. 2023.

\bibitem{thomas2023risOTFS}
A.~Thomas, K.~Deka, S.~Sharma, and N.~Rajamohan, ``{IRS}-assisted {OTFS} system: {D}esign and analysis,'' {\em IEEE Trans. Veh. Technol.}, vol.~72, pp.~3345--3358, Mar. 2023.

\bibitem{raviteja2018interference}
P.~Raviteja, K.~T. Phan, Y.~Hong, and E.~Viterbo, ``Interference cancellation and iterative detection for orthogonal time frequency space modulation,'' {\em IEEE Trans. Wireless Commun.}, vol.~17, pp.~6501--6515, Oct. 2018.

\bibitem{JMLR:v11:journee10a}
M.~Journ{{\'e}}e, Y.~Nesterov, P.~Richt{{\'a}}rik, and R.~Sepulchre, ``Generalized power method for sparse principal component analysis,'' {\em J. Mach. Learn. Res.}, vol.~11, pp.~517--553, Dec. 2010.

\bibitem{fischer2005precoding}
R.~F. Fischer, {\em Precoding and Signal Shaping for Digital Transmission}.
\newblock New York, NY, USA: John Wiley \& Sons, 2002.

\bibitem{dinan2023ris}
M.~H. Dinan, M.~D. Renzo, and M.~F. Flanagan, ``{RIS}-assisted receive quadrature spatial modulation with low-complexity greedy detection,'' {\em IEEE Trans. Commun.}, vol.~71, pp.~6546--6560, Nov. 2023.

\bibitem{etsiTR36913}
\emph{{5G}; {S}tudy on channel model for frequencies from 0.5 to 100 {GHz}}, document 3GPP TR 38.901, version 17.0.0, Release 17, European Telecommunications Standards Institute, Apr. 2022.

\end{thebibliography}

\end{document}